\documentclass[%
 reprint,
superscriptaddress,
 amsmath,amssymb,
 aps,
 prl,
floatfix,
]{revtex4-2}

\usepackage{graphicx}% Include figure files
\usepackage{dcolumn}% Align table columns on decimal point
\usepackage{bm}% bold math
\usepackage[]{hyperref}% add hypertext capabilities
\usepackage[mathlines]{lineno}% Enable numbering of text and display math
\usepackage{xcolor} % XXX own package Ida
\usepackage{upgreek} % XXX own package Ida
\usepackage[utf8]{inputenc}
\usepackage[]{babel}
\UseRawInputEncoding % for Arxiv
\begin{document}

\preprint{APS/123-QED}

\title{Electron Heat Flux and Whistler Instability in the Earth's Magnetosheath} 

\author{Ida Svenningsson}
\affiliation{
Swedish Institute of Space Physics, Uppsala 751 21, Sweden
}%
\affiliation{
Department of Physics and Astronomy, Uppsala University, Uppsala 751 20, Sweden
}%
\affiliation{
Department of Physics, Chalmers University of Technology, Göteborg 412 96, Sweden
}%

\author{Emiliya Yordanova}
\affiliation{
Swedish Institute of Space Physics, Uppsala 751 21, Sweden
}%

\author{Yuri V. Khotyaintsev}
\affiliation{
Swedish Institute of Space Physics, Uppsala 751 21, Sweden
}%
\affiliation{
Department of Physics and Astronomy, Uppsala University, Uppsala 751 20, Sweden
}%

\author{Mats Andr\'e}
\affiliation{
Swedish Institute of Space Physics,  Uppsala 751 21, Sweden
}%

\author{Giulia Cozzani}
\affiliation{
Laboratory of Physics and Chemistry of the Environment and Space (LPC2E), OSUC, Univ Orleans, CNRS, CNES, 45071 Orleans, France 
}%

\author{Alexandros Chasapis}
\affiliation{Laboratory for Atmospheric and Space Physics, University of Colorado, Boulder, Colorado 80303, USA}

\author{Steven J. Schwartz}
\affiliation{Laboratory for Atmospheric and Space Physics, University of Colorado, Boulder, Colorado 80303, USA}
\affiliation{Department of Physics, Imperial College London, London, SW7 2AZ, UK}

\date{\today}

\begin{abstract} % max 600 characters incl. spaces
Despite heat flux's role in regulating energy conversion in collisionless plasmas, its properties and evolution in the magnetosheath downstream of the Earth's bow shock are scarcely explored. We use MMS \textit{in situ} measurements to quantify and characterize the electron heat flux in the magnetosheath. We find that the heat flux is shaped by the magnetosheath magnetic field as it drapes around the magnetosphere. While it is affected by solar wind upstream conditions and increases with magnetic field strength, it is not substantially changed by local magnetosheath processes. Also, the heat flux is limited by whistler instability thresholds.

\end{abstract}

\maketitle

\emph{Introduction} -
Electrons control the thermal energy transfer in plasmas due to their low mass compared to ions \cite{verscharen_case_2022}, making the electron heat flux central to the total energy budget.
In the weakly collisional regime, the electron heat flux is suppressed below the collisional Spitzer-H{\"a}rm \cite{spitzer_transport_1953} prediction, as observed in the solar wind \cite{bale_electron_2013} and the intracluster medium of galaxy clusters \cite{binney_x-ray_1981,markevitch_chandra_2003}.
Instead, wave-particle interactions are believed to suppress the heat flux on ion and electron kinetic scales in the heliospheric \cite{scime_regulation_1994} and astrophysical plasmas \cite{roberg-clark_wave_2018}. 

The electron heat flux $\bm{q}_{e}$ is defined as the third-order moment of the electron velocity distribution function (eVDF):
\begin{equation}
    \bm{q}_{e} = \frac{m_e}{2}\int f_{e}(\bm{v}) 
    |\bm{w}|^2\bm{w}
    \,{d}^3 v,
    \label{eq:q_def}
\end{equation}
where $f_e(\bm{v})$ is the eVDF, $\bm{w}=\bm{v}-\bm{V}_{e}$ is the velocity $\bm{v}$ in the electron bulk flow ($\bm{V}_e$) frame, and $m_e$ the electron mass.
The heat flux can be calculated directly from the eVDF in spacecraft observations \cite{scime_regulation_1994,bale_electron_2013,halekas_electron_2021,coburn_regulation_2024,gershman_evolution_2024}.
However, since $\bm{q}_{e}$ is sensitive to noise related to low count rates, detailed calculations of $\bm{q}_{e}$ from spacecraft data have mostly been done for case studies \cite{coburn_regulation_2024,gershman_evolution_2024}.
In the solar wind (SW), $\bm{q}_{e}$ can be obtained by fitting the eVDF core, halo, and strahl components as drifting bi-Maxwellians \cite{gary_whistler_1994, gary_electron_1999}.
In hybrid simulations, $\bm{q}_{e}$ can be inferred through thermodynamic closures such as Landau-fluid (LF) type closures \cite{finelli_bridging_2021,sulem_landau_2015}. Data-driven closures using machine learning \cite{huang_machine-learning_2025} and sparse regression 
\cite{ingelsten_data-driven_2025} have also been suggested.
However, it remains a challenge to estimate $\bm{q}_{e}$ in spacecraft data, notably when the eVDF is strongly non-Maxwellian \cite{graham_nonmaxwellianity_2021}.

Collisionless shocks are associated with strong particle acceleration and heating in various astrophysical environments such as supernova remnants \cite{pohl_electron_1998}, interplanetary shocks \cite{kahler_solar_2007,yang_strongest_2018}, and planetary bow shocks \cite{phillips_ulysses_1993,andreone_properties_2022}.
Among these, the Earth's bow shock (BS) is uniquely accessible to \textit{in situ} measurements of electron dynamics.
At the BS, electrons are accelerated and heated adiabatically, creating separated hot and cold components. Additionally, non-adiabatic effects due to, e.g., wave-particle interactions \cite{amano_statistical_2024} become significant at high Mach numbers in the de Hoffmann‐Teller (HT) reference frame \cite{lalti_adiabatic_2024}.
This heating gives the eVDF in the downstream magnetosheath (MSH) region a 'flat top' shape at low energies \cite{feldman_electron_1983}; significantly different from the eVDF in the SW.
The heating is also expected to change $\bm{q}_{e}$, but the resulting change in $\bm{q}_{e}$ across the BS is not known.
Moreover, $\bm{q}_{e}$ may be altered locally in the MSH by reconnecting current sheets \cite{retino_situ_2007,yordanova_electron_2016,voros_mms_2017,phan_electron_2018,stawarz_properties_2019,stawarz_turbulence-driven_2022}, wave-particle interactions \cite{svenningsson_kinetic_2022,svenningsson_whistler_2024,chen_evidence_2019,afshari_direct_2024,breuillard_properties_2018}, or magnetic reconnection at the magnetopause (MP) \cite{gonzalez_magnetic_2016,gershman_evolution_2024}.
The relevant space mission allowing the access to electron scales is the Magnetospheric Multiscale (MMS) mission \cite{burch_magnetospheric_2016}, equipped with state-of-the-art instrumentation, permitting the estimation of higher-order moments of the eVDF, such as the heat flux.

In the SW, $\bm{q}_{e}$ can be regulated by different wave instabilities, of which the whistler wave dominates for $\beta_{e\parallel}\gtrsim0.1$ \cite{gary_whistler_1994}, where $\beta_{e\parallel}=2\mu_0n_eT_{e\parallel}/B^2$ and $n_e$ and $T_{e\parallel}$ are the electron density and temperature parallel to the magnetic field $\bm{B}$, and $B = |\bm{B}|$.
Heat flux instabilities can generate both parallel-propagating \cite{gary_whistler_1994} and highly oblique \cite{vasko_whistler_2019} whistler waves, the interplay of which results in a heat flux reduction \cite{micera_particle--cell_2020}. However, the regulation of $\bm{q}_{e}$ in the MSH, a compressible, high-beta region in contrast to the SW \cite{huang_existence_2017,yordanova_current_2020}, remains unknown.
Previous analyses also assumed a SW-type eVDF. Studying the MSH allows us to investigate the more general applicability of the scenario where the heat flux is regulated by these instabilities.

Whistler waves are regularly observed in the MSH. While the most commonly proposed generation mechanism requires a high electron temperature anisotropy \cite{gary_whistler_1996}, whistler waves are not typically associated with the high electron temperature anisotropy required to trigger waves \cite{svenningsson_whistler_2024}. This can be partly explained by highly non-Maxwellian eVDFs \cite{kitamura_observations_2020,page_generation_2021,svenningsson_kinetic_2022}.
Therefore, heat flux instabilities should be investigated as alternative candidates for whistler wave generation. 

To explore the poorly understood nature of electron heat flux downstream of collisionless shocks, we use measurements from the MMS mission to statistically study the electron heat flux in the MSH and its evolution from the BS to the MP. We also explore the effect of local heat flux regulation by investigating the relationship with whistler waves.

\begin{figure}[!t]
    \centering
    \includegraphics[width=\linewidth]{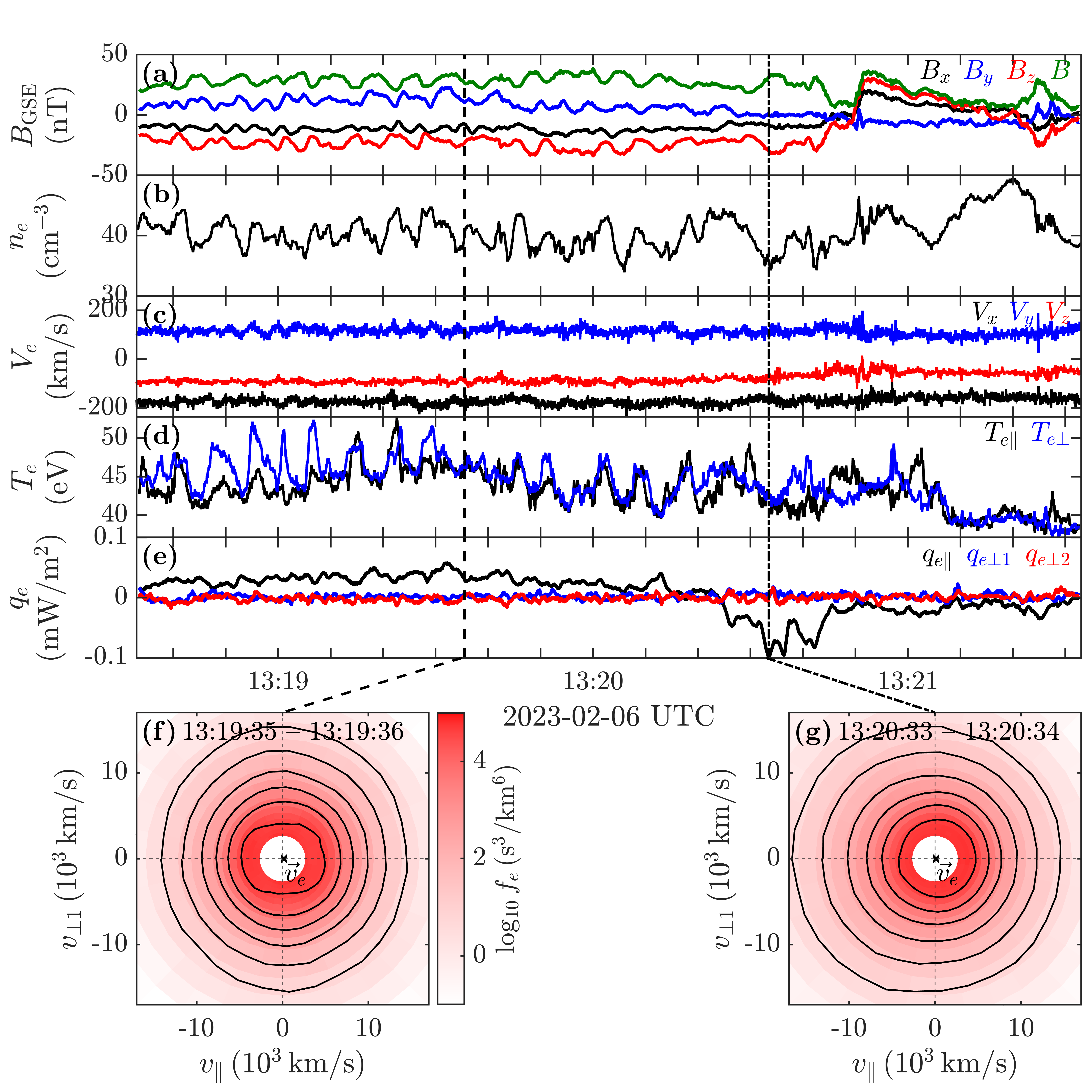}
    \caption{Three-minute interval from the MSH.
    MMS measurements of the magnetic field $\boldsymbol{B}$ (a) and the electron moments: 
    (b) density $n_e$,
    (c) bulk velocity $\bm{V}_e$,
    (d) temperatures parallel ($T_{e\parallel}$) and perpendicular ($T_{e\perp}$) to $\boldsymbol{B}$,
    (e) heat flux $\boldsymbol{q}_e$ in field-aligned components.
    % In (a) and (e), we have overplotted data from MMS1-3.
    (f) and (g) show 2D eVDFs $f_e(v_\parallel,v_{\perp1})$ (averaged over 1 second) taken from 13:19:35 and 13:20:33, respectively,
    where $v_\parallel=\bm{v}\cdot \bm{\hat{b}}$ and $v_{\perp1}=\bm{v}\cdot (\bm{\hat{b}}\times(\bm{\hat{e}}\times\bm{\hat{b}}))$, and $\bm{\hat{b}}=\bm{B}/B$ and $\bm{\hat{e}}=\bm{E}/E$ are the magnetic and electric field directions, respectively.
    }
    % plot done in project4_heatflux_paperplots.m (is = 2; it = 7;)
    \label{fig:heatflux_example}
\end{figure}

\emph{Data and methods} - 
We use MMS data from the 2023 \textit{unbiased magnetosheath campaign} \cite{svenningsson_kinetic_2022}, where 3 minutes of burst-mode data were sampled every 9 minutes providing 14 full inbound MSH crossings with a total of $13.9\,\rm hours$ of data.
The fluxgate magnetometer (FGM) \cite{russell_magnetospheric_2016} provides the magnetic field vector $\boldsymbol{B}$ and the Fast Plasma Investigation (FPI) \cite{pollock_fast_2016} instruments give the full 3D eVDF every $30\,\rm ms$.
We use the highest-resolution burst-mode data to avoid effects of lossy compression of fast-mode data \cite{barrie_wavelet_2019}.
We pre-process the electron heat flux to remove non-physical noise originating from low-counting statistics. The noise removal algorithm, described in the Supplementary material, involves removing extreme values, correcting for the spacecraft spin tone, and smoothing the signal with a 1-second window.

\emph{Results} - 
Figure~\ref{fig:heatflux_example} shows one of the 3-minute intervals from the MSH. Panels (a)-(d) show $\boldsymbol{B}$, $n_e$, $\boldsymbol{V}_e$, and parallel and perpendicular temperatures ($T_{e\parallel}$, $T_{e\perp}$ with respect to $\boldsymbol{B}$), respectively.
In the first half of the interval, $B$ and $n_e$ display an anti-correlation typical of mirror-mode waves common in the quasi-perpendicular MSH configuration \cite{dimmock_statistical_2015}, where the mirror modes modulate $T_{e\perp}$ \cite{yao_electron_2018}. 
Figure~\ref{fig:heatflux_example}e shows the electron heat flux $\bm{q}_e$ in field-aligned coordinates. The parallel component $q_{e\parallel}$ dominates, consistent with approximately gyrotropic observed eVDFs. We note that $\boldsymbol{q}_e$ changes direction in the vicinity of a sharp $\bm{B}$ variation at 13:20:50. 
Figure~\ref{fig:heatflux_example}f-g show the 2D eVDFs $f_e(v_\parallel,v_{\perp1})$ at two times: at 13:19:35, there is a parallel heat flux, and at 13:20:33, the heat flux is anti-parallel. The associated asymmetry is visible in $f_e$, especially in panel (g) where $f_e$ is skewed towards negative $v_\parallel$ at high energies.
Thus, the heat flux can vary locally in the MSH, and its behavior is not directly predictable from other moments ($n_e$, $\boldsymbol{V}_e$, $T_{e\parallel}$, and $T_{e\perp}$).

\begin{figure}[!t]
    \centering
    \includegraphics[width=\linewidth]{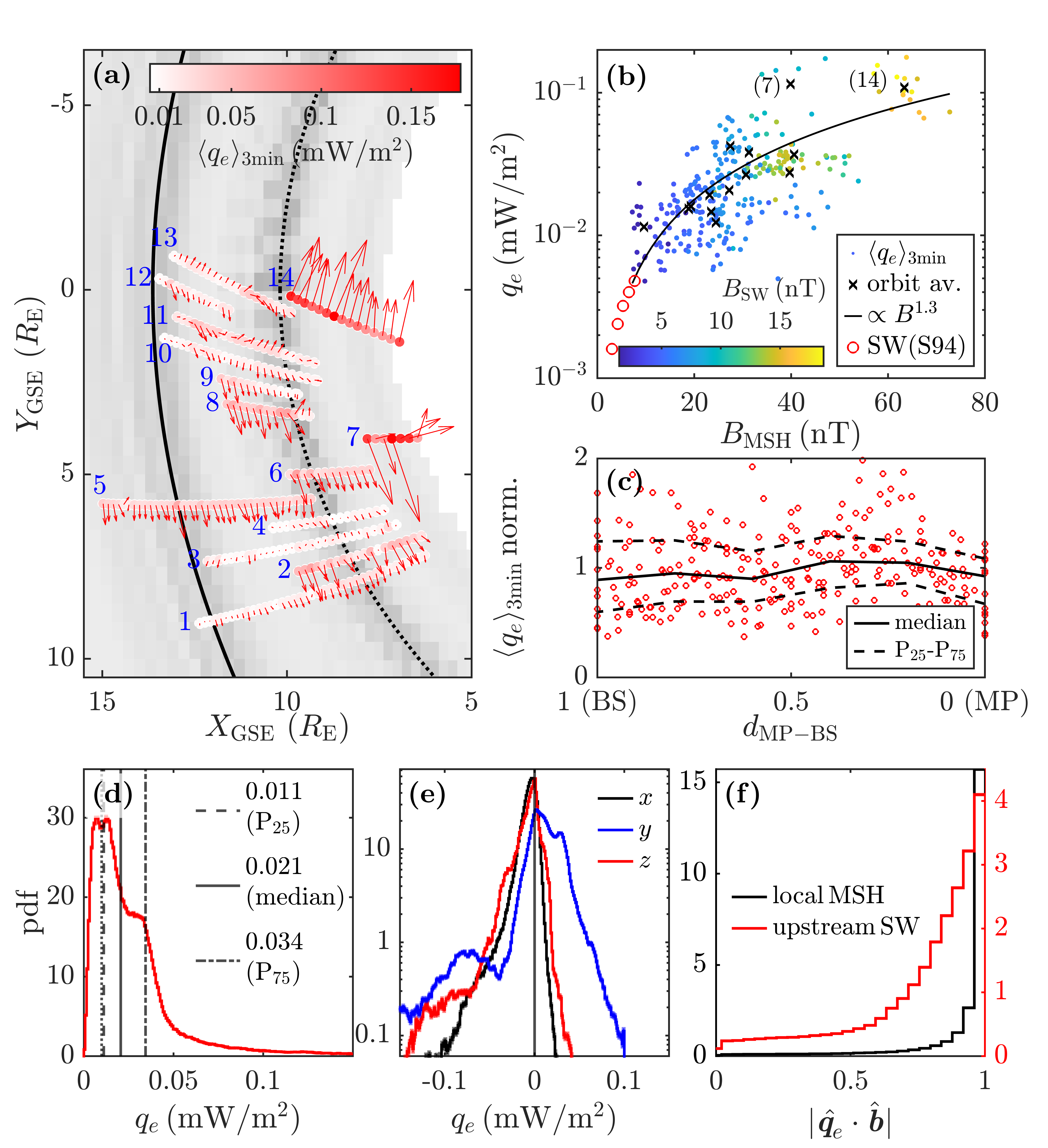}
    \caption{Electron heat flux across the MSH.
    (a) MMS location in each 3-minute burst interval in the GSE $XY$ plane ($Z_{\rm GSE}{\sim}(-7,-5)R_{\rm E}$). Color indicates $\langle q_{e}\rangle_{\rm 3min}$ and arrows the projected $\langle\bm{q}_e\rangle_{\rm 3min}$.
    Orbits are labeled chronologically 1-14 with blue numbers.
    (b) $\langle q_{e}\rangle_{\rm 3min}$ versus $B_{\rm MSH}$; the color scale shows $B_{\rm SW}$. Black curve shows the fit $q_e\approx 3\times10^{-4}B^{1.3}$ based on the orbit-average values (black crosses). Red circles show approximate values of $B$ and $q_e$ from the SW, obtained by reading off the axes in Fig.~13 in Scime+~1994 (S94) \cite{scime_regulation_1994}.
    (c) Evolution of $\langle q_{e}\rangle_{\rm 3min}$ (normalized to the mean in each orbit) with the fractional distance $d_{\rm MP-BS}$. Solid curve shows the median in bins of width 0.2, and dashed curves show the $25^{\rm th}$ and $75^{\rm th}$ percentiles in the bins.
    Individual orbits are shown in red.
    (d) Overall $q_{e}$ in the MSH; median value in solid, $25^{\rm th}$ and $75^{\rm th}$ percentiles in dashed and dash-dotted. Dotted line indicates the estimated uncertainty $0.01\,\rm mW/m^2$ (see Supplementary material).
    (e) GSE components of $\bm{q}_e$.
    (f) alignment of $\hat{\bm{q}}_e=\bm{q}_e/q_e$ with the local magnetic field direction $\hat{\bm b}_{\rm MSH}=\bm{B}_{\rm MSH}/B_{\rm MSH}$ (black) and the upstream solar wind magnetic field direction $\hat{\bm b}_{\rm SW}=\bm{B}_{\rm SW}/B_{\rm SW}$ (red).
    }
    % plot done in project4_heatflux_paperplots_new.m
    \label{fig:heatflux_map}
\end{figure}

We now investigate whether local variations in the heat flux such as the one shown in Figure~\ref{fig:heatflux_example}e accumulate to a significant change across the MSH. 
We quantify the large-scale heat flux with $\langle q_{e}\rangle_{\rm 3min}$, the mean magnitude of $\bm{q}_{e}$ in each 3-minute burst interval.
Figure~\ref{fig:heatflux_map}a shows the locations of MMS in the  geocentric solar ecliptic (GSE) $XY$ plane with color and arrows indicating $\langle q_{e}\rangle_{\rm 3min}$ and the mean vectors $\langle \bm{q}_{e}\rangle_{\rm 3min}$, respectively.
The solid and dotted curves show models of the BS and MP, respectively, based on average SW conditions (retrieved from the OMNI database \cite{OMNI}) during the considered intervals, and the variation of their locations is indicated in gray (see details in Ref.~\cite{svenningsson_classifying_2025}).
The $\bm{q}_e$ direction is tied to the magnetic field lines which are approximately aligned with the BS surface since the MSH is observed in a quasi-perpendicular configuration \cite{svenningsson_whistler_2024}.
In each orbit, $\langle q_{e}\rangle_{\rm 3min}$ does not change within the MSH with a few exceptions (e.g., $\langle q_{e}\rangle_{\rm 3min}$ increasing with BS distance in orbit \#1 and the rotation in orbit \#7).
The SW conditions, such as the interplanetary magnetic field direction $\hat{\bm b}_{\rm SW}$, are mostly stable during the individual orbits while in orbits where $\hat{\bm b}_{\rm SW}$ rotates, this is accompanied with rotation in $\bm{q}_e$ (see Supplementary material).
For two orbits \#7 and \#14 the MSH is highly compressed, with MSH measurements well within the average MP. This is due to a high-speed stream (orbit \#7) and a coronal mass ejection (orbit \#14). In both cases, $\langle q_{e}\rangle_{\rm 3min}$ is 3-10 times higher than in the other orbits.

Figure~\ref{fig:heatflux_map}b shows that $\langle q_{e}\rangle_{\rm 3min}$ increases with the local MSH magnetic field strength $B_{\rm MSH}$.
The same dependence is seen with respect to $B_{\rm SW}$ (color scale) and with the SW dynamic pressure (see Supplementary material).
The fit (black curve) based on the orbit-averages reveals that $q_e$ is approximately proportional to $B_{\rm MSH}$.
Interestingly, this scaling with $B$ is similar to previous observations in the SW at $1{-}5\,\rm AU$ (red circles, S94 \cite{scime_regulation_1994}).

In Figure~\ref{fig:heatflux_map}c, we study the $\bm{q}_e$ evolution across the MSH using $\langle q_{e}\rangle_{\rm 3min}$ normalized to the mean value in each orbit.
We estimate the location relative to the MP and the BS  with the fractional distance $d_{\rm MP-BS}$, defined so that $d_{\rm MP-BS}=0$ at the MP and $d_{\rm MP-BS}=1$ at the BS \cite{dimmock_statistical_2013,svenningsson_classifying_2025}.
We show the individual orbits in red and the black curves show the median (solid), $25^{\rm th}$ and $75^{\rm th}$ percentiles (dashed) in bins of width $d_{\rm MP-BS}=0.2$.
While there is a large variation across different orbits, there is no clear evolution of the normalized $\langle q_{e}\rangle_{\rm 3min}$ with $d_{\rm MP-BS}$, indicating that the electron heat flux is not significantly created or destroyed in the MSH.

Figure~\ref{fig:heatflux_map}d-f shows histograms of $\bm{q}_e$ without time averaging, i.e., obtained from the entire dataset, corresponding to $1{,}671{,}203$ eVDFs. The median $q_{e}$ is $0.021\,\rm mW/m^2$ and $50\,\%$ of the data is within $0.012$ and $0.035\,\rm mW/m^2$ (see Figure~\ref{fig:heatflux_map}d).
This is slightly higher than reported values of $0.005{-}0.01\,\rm mW/m^2$ in the SW at $1\,\rm AU$ \cite{scime_regulation_1994,cattell_narrowband_2020,salem_precision_2023}. % Median 0.01 for coherent and 0.0077 for incoherent whistler waves. STEREO observations; Salem2023: 0.005, slightly lower in fast SW
The $x$ component is negative ($q_{ex}<0$; black in Figure~\ref{fig:heatflux_map}e) similar to the anti-sunward strahl in the SW.
Its southward direction ($q_{ez}<0$; red curve) indicates that the heat flux is directed around the magnetosphere since MMS is located at $Z_{\rm GSE}<0$ during the entire dataset (not shown).
As expected for approximately gyrotropic distributions, $\bm{q}_e$ is aligned with the local magnetic field ($|\hat{\bm q}_e\cdot\hat{\bm b}_{\rm MSH}|\approx 1$, black in Figure~\ref{fig:heatflux_map}f).
Moreover, since $|\hat{\bm q}_e\cdot\hat{\bm b}_{\rm SW}|$ is predominantly 1 (Figure~\ref{fig:heatflux_map}f) the heat flux likely follows the magnetic field draping around the magnetosphere (see also Supplementary material).

\begin{figure}[!t]
    \centering
    \includegraphics[width=\linewidth]{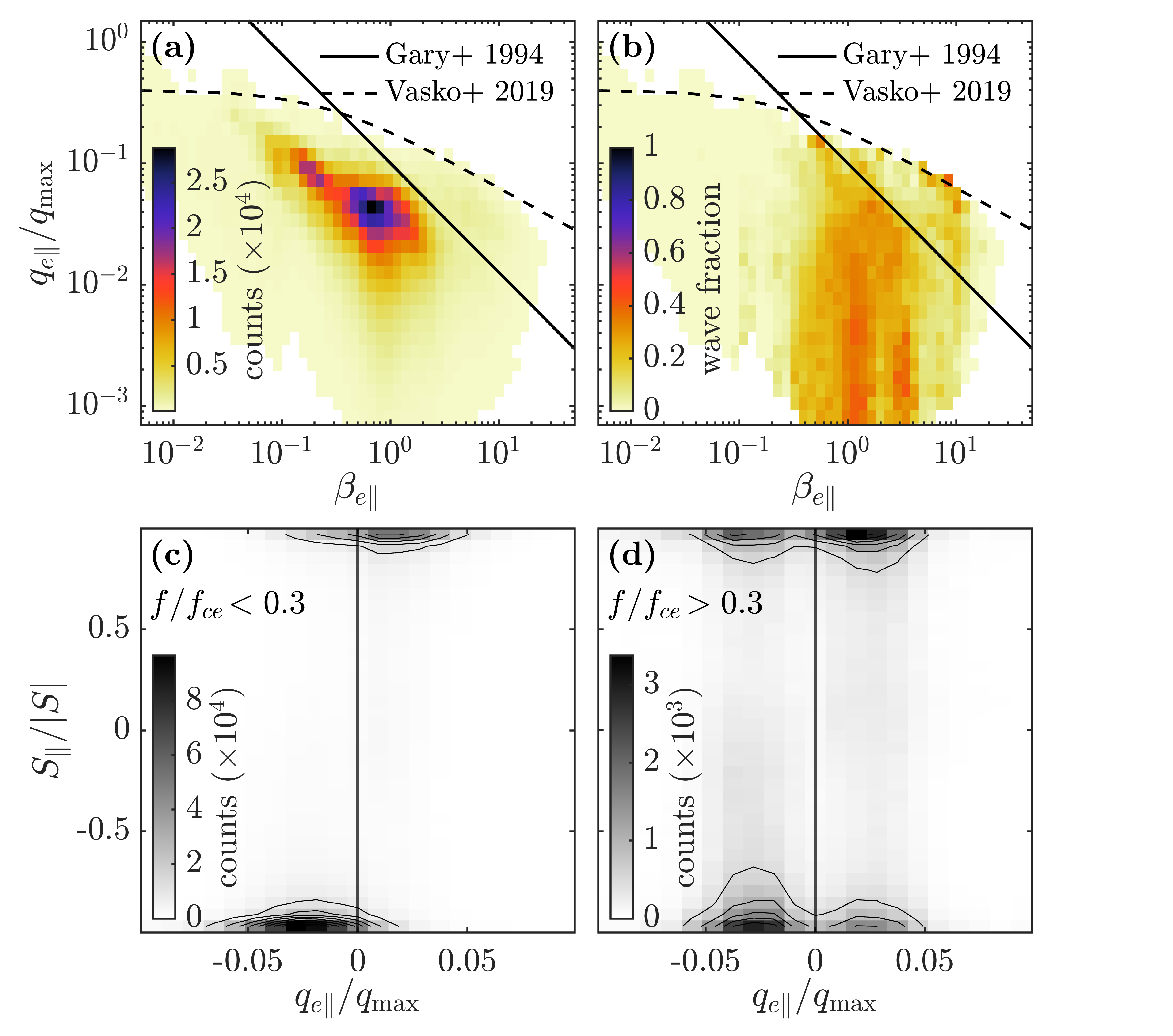}
    \caption{Whistler heat flux instabilities.
    (a) 2D histogram of $|q_{e\parallel}|/q_{\rm max}$ and $\beta_{e\parallel}$ with each count corresponding to one $30\,\rm ms$ electron measurement.
    (b) Whistler wave occurrence in the $\beta_{e\parallel}-|q_{e\parallel}|/q_{\rm max}$ parameter space. 
    We show two whistler instability thresholds: the parallel heat flux instability (solid: Gary+~1994 \cite{gary_whistler_1994}: $|q_{e\parallel}|/q_{\rm max}=0.10\beta_{e\parallel}^{-0.90}$ for a growth rate $\gamma=10^{-3}\Omega_p$, where $\Omega_p$ is the angular proton cyclotron frequency)
    and the oblique heat flux fan instability 
    (dashed: Vasko+~2019 \cite{vasko_whistler_2019}: $|q_{e\parallel}|/q_{\rm max}=0.20\left[\beta_{e\parallel}+0.25\right]^{-0.50}$). % threshold for no strahl velocity...
    (c) Normalized Poynting flux $S_\parallel/|S|$ of low-frequency whistler waves ($f/f_{{\rm c}e}<0.3$) versus $q_{e\parallel}/q_{\rm max}$.
    (d) Same as (c) but for high-frequency ($f/f_{{\rm c}e}>0.3$) whistler waves.
    }
% plot done in project4_heatflux_paperplots_new.m    
\label{fig:heatflux_whistlers}
\end{figure}

To understand how the heat flux is regulated in the magnetosheath, we compare the heat flux distribution to instability thresholds for the whistler heat flux instability, which is the most likely instability since $\beta_{e\parallel}\gtrsim0.1$ in our dataset \cite{gary_whistler_1994}.
Figure~\ref{fig:heatflux_whistlers}a shows the $\beta_{e\parallel}-|q_{e\parallel}|/q_{\rm max}$ parameter space, where $q_{\rm max}=\frac{3}{2}m_en_ev_{\rm th}^3$ is the free-streaming heat flux \cite{bale_electron_2013} and $v_{\rm th}=\sqrt{T_e/m_e}$ is the thermal speed.
We show the instability threshold for parallel propagating waves (solid) \cite{gary_whistler_1994} and oblique waves (dashed) \cite{vasko_whistler_2019}, 
The bulk of the data is bounded by the parallel threshold and the oblique threshold provides an upper boundary on the distribution. This suggests that the whistler instability regulates the heat flux in the MSH.

To further explore the role of whistler instabilities, we compare the heat flux to whistler wave packets analyzed previously in Ref.~\cite{svenningsson_whistler_2024}.
The wave packets are mainly found parallel-propagating with respect to $\bm B$ and appear in two frequency bands; the majority with $f/f_{{\rm c}e}\in(0.1,0.3)$ and a smaller population with $f/f_{{\rm c}e}\in(0.3,0.6)$, where $f_{{\rm c}e}=eB/(2\pi m_e)$ is the electron cyclotron frequency \cite{svenningsson_whistler_2024}.
In Figure~\ref{fig:heatflux_whistlers}b we show the fraction of data points in the MSH during which whistler wave packets are observed. Whistler activity is mainly observed for $\beta_{e\parallel}>0.1$ as noted in Ref.~\cite{svenningsson_whistler_2024}. Apart from an increased wave activity at the oblique instability threshold (dashed) for $\beta_{e\parallel}\approx10$, whistler activity does not correlate with a high heat flux close to the instability thresholds.
However, the whistler wave distribution follows the parallel whistler instability threshold (solid), suggesting that there is a connection between heat flux and whistler waves.

In general, due to the asymmetry of the eVDF, whistler waves produced by the heat flux instability propagate along the magnetic field in the same direction as the heat flux \cite{verscharen_electron-driven_2022}, while the temperature anisotropy instability generates both parallel and anti-parallel waves.
Therefore, we investigate the wave propagation direction using the Poynting flux $\bm{S}$, which is parallel to the group velocity.
Figure~\ref{fig:heatflux_whistlers}c shows $S_\parallel/|\bm{S}|$, the normalized parallel Poynting flux with respect to $\bm B$, versus $q_{e\parallel}/q_{\rm max}$, for low-frequency ($f/f_{{\rm c}e}<0.3$) whistler wave packets, i.e, where the majority of the observed waves are found. The clustering in the top right and bottom left corners indicates that the wave packets are predominantly aligned with the heat flux, consistent with the whistler heat flux instability.
On the other hand, the propagation of high-frequency waves ($f/f_{{\rm c}e}>0.3$; Figure~\ref{fig:heatflux_whistlers}d) has less alignment % does not display the same alignment 
with $q_{e\parallel}$.

\emph{Discussion} - % \label{sec:discussion}
In this Letter, we statistically analyze the electron heat flux across the MSH, from the BS to the MP.
We find that the heat flux in the MSH shares traits of the heat flux in the upstream SW, as it is mainly field-aligned, has an anti-sunward component. Some properties are altered; most notably, the direction follows the field line draping around the magnetosphere and the magnitude is higher than typical values in the solar wind, especially when the magnetic field is strong.
Due to the order-of-magnitude variations between different orbits, the upstream conditions likely impact the heat flux downstream of the bow shock. Importantly, both cases of highly compressed MSH show significantly higher values compared to other orbits.

We find that $q_e$ increases proportionally to the local magnetic field strength $B_{\rm MSH}$ and the upstream $B_{\rm SW}$. 
A correlation between $q_e$ and $B_{\rm SW}$ has been observed in the SW at $1{-}5\,\rm AU$ \cite{scime_regulation_1994}; however, this was also a radial effect as both $q_e$ and $B$ decrease with distance from the sun.
In this work, the reason for the proportionality is not clear.
Possibly the correlation exists in the SW even at 1~AU.
Moreover, considering the high $q_e$ in the MSH compared to the SW, it seems that $q_e$ increases at the BS. With the present dataset, we cannot determine the relative impacts of upstream conditions and shock acceleration processes.
This requires larger datasets as well as continuous $\bm{q}_e$ measurements from the SW to the MSH.

% STEADY IN THE SHEATH
The magnitude of the electron heat flux remains relatively steady within the MSH during individual orbits. Although it can vary locally, the cumulative effect of potential sources and sinks does not globally affect the heat flux magnitude.
% BOUNDARIES
One potential source for heat flux is magnetic reconnection at the MP \cite{gershman_evolution_2024}. However, we do not observe an increase of heat flux close to this boundary ($d_{\rm MP-BS}\approx0$). This suggests that the heat flux created at the MP remains in the vicinity of the MP. 

% WHISTLER
The heat flux is limited by the parallel \cite{gary_whistler_1994} and oblique \cite{vasko_whistler_2019} whistler instability thresholds.
Since the MSH whistlers tend to propagate quasi-parallel to $\bm{B}$ \cite{svenningsson_whistler_2024}, the parallel instability is expected to dominate. However, given the upper bound by the oblique threshold, there may be impact from both instabilities.
We find that there is no increase in whistler wave activity when the heat flux is close to the instability thresholds, indicating that the heat flux is not the major source of waves. As the same instabilities bound the heat flux in the SW \cite{gary_whistler_1994,cattell_parker_2022,halekas_electron_2021}, the fact that the heat flux is constrained by these instability thresholds can be a remainder of SW dynamics, where the heat flux has already been actively regulated by the whistler instability.

On the other hand, a possible whistler-heat flux instability is supported by the alignment between low-frequency ($f/f_{{\rm c}e}<0.3$) whistlers and the heat flux direction. Also, the wave occurrence is constrained by the parallel whistler instability, similarly to what has been observed in the SW \cite{cattell_parker_2022}. Thus, this instability cannot be out-ruled.
One possibility is that the asymmetric features typical of the heat flux instability exist in the whistler-resonant parts of the eVDF (satisfying the first-order cyclotron resonance condition, typically a few times the thermal energy \cite{svenningsson_whistler_2024}).
Thus, the heat flux calculated from the entire distribution may not necessarily predict an instability \cite{page_generation_2021}. 
Also, high (low) electron temperature anisotropies can trigger (quench) the onset of the whistler heat flux instability \cite{tong_whistler_2019,tong_statistical_2019}, which could affect the spread in wave occurrence.
Considering the agreement with instability thresholds and the directional alignment, we suggest that the heat flux whistler instabilities play a role in the MSH.

For a reliable estimation of the electron heat flux, the analysis presented here required removing noise, non-physical data spikes, and correcting for a spin signal (see Supplementary material). Previously, the heat flux has been obtained from the eVDF with another method \cite{gershman_evolution_2024}, by restricting the energy range and estimating the errors from Poisson statistics. However, this is more suitable for case studies due to the heavy processing of the full eVDF at each time step. Our approach provides an alternative better suited for statistical studies when a large amount of data needs to be processed. 

\emph{Conclusions} - % \label{sec:summary}
The electron heat flux is important for understanding energy conversion in space, astrophysical and laboratory plasma. Using 13.9 hours of burst-mode data from the Magnetospheric Multiscale (MMS) mission, we statistically study the electron heat flux downstream of the Earth's bow shock, how it evolves inside the magnetosheath, and its potential regulation by whistler instabilities. We find that the magnetosheath electron heat flux is slightly higher than that in the solar wind at 1~AU and that it drapes around the magnetosphere following the magnetic field lines. Its magnitude varies with upstream solar wind conditions and increases with magnetic field strength. However, local processes in the magnetosheath do not substantially change the heat flux during propagation from the bow shock to the magnetopause. The heat flux is constrained by the parallel and oblique whistler instability thresholds, and the alignment of low-frequency whistler wave propagation with the heat flux direction further supports a connection between electron heat flux and whistler waves.

Being the first statistical study of electron heat flux downstream of the Earth's bow shock, our findings should be instructive when analyzing heat conduction downstream of astrophysical shocks such as supernova remnants \cite{caprioli_simulations_2014,cui_two-temperature_1992,inoue_turbulence_2009}, as well as other collisionless, high-beta plasmas which are not accessible for \textit{in situ} measurements, including the interstellar medium \cite{lee_turbulence_2020} and black hole accretion disks \cite{ressler_electron_2015,event_horizon_telescope_collaboration_first_2022}.
Our results thus provide a valuable contribution to the  understanding the global energy budget in the MSH and can also serve as a reference for modelling the electron heat flux in high-beta, compressible plasmas.

\begin{acknowledgments}
\emph{Acknowledgments} - 
MMS data are available at the MMS Science Data Center \footnote{See \url{https://lasp.colorado.edu/mms/sdc/public}.}.
Data analysis was performed using the \verb+irfu-matlab+ package \footnote{See \url{https://github.com/irfu/irfu-matlab}.}.
We thank the MMS team and instrument PIs for data access and support.
We thank Heli Hietala and Istvan Pusztai for valuable input.
We acknowledge support from the Swedish Research Council Grant 2016‐0550 and the Swedish National Space Agency Grants
158/16 and 192/20.
This research was supported by the International Space Science Institute (ISSI) in Bern, through the ISSI International Team project \#23-588 ('Unveiling Energy Conversion and Dissipation in Non-Equilibrium Space Plasmas').
The work was supported by the Knut and Alice Wallenberg foundation. 
The work of GC is supported by the Integration Fellowship of Le Studium Loire Valley Institute for Advanced Studies.
\end{acknowledgments}

%  ö {\"o} Ä {\"A} Å {\AA} 

\bibliographystyle{apsrev4-2}
\bibliography{main}% Produces the bibliography via BibTeX.

@article{OMNI,
    author = {Papitashvili, Natalia E. and King, Joseph H.},
    journal = {{NASA Space Physics Data Facility}},
    title = {OMNI Hourly Data},
    doi = {https://doi.org/10.48322/1shr-ht18},
    year={2020}
}

@article{markevitch_chandra_2003,
doi = {10.1086/374656},
url = {https://doi.org/10.1086/374656},
year = {2003},
month = {feb},
publisher = {},
volume = {586},
number = {1},
pages = {L19},
author = {Markevitch, M. and Mazzotta, P. and Vikhlinin, A. and Burke, D. and Butt, Y. and David, L. and Donnelly, H. and Forman, W. R. and Harris, D. and Kim, D.-W. and Virani, S. and Vrtilek, J.},
title = {Chandra Temperature Map of A754 and Constraints on Thermal Conduction},
journal = {The Astrophysical Journal},
}

@article{binney_x-ray_1981,
	title = {X-ray emission from {M87} - {A} pressure confined cooling atmosphere surrounding a low mass galaxy},
	volume = {247},
	issn = {0004-637X, 1538-4357},
	url = {http://adsabs.harvard.edu/doi/10.1086/159055},
	doi = {10.1086/159055},
	language = {en},
	urldate = {2026-02-16},
	journal = {The Astrophysical Journal},
	author = {Binney, J. and Cowie, L. L.},
	month = jul,
	year = {1981},
	pages = {464},
}

@article{cui_two-temperature_1992,
	title = {Two-{Temperature} {Models} of {Old} {Supernova} {Remnants} with {Ion} and {Electron} {Thermal} {Conduction}},
	volume = {401},
	issn = {0004-637X},
	url = {https://ui.adsabs.harvard.edu/abs/1992ApJ...401..206C},
	doi = {10.1086/172053},
	urldate = {2026-02-16},
	journal = {The Astrophysical Journal},
	publisher = {IOP},
	author = {Cui, Wei and Cox, Donald P.},
	month = dec,
	year = {1992},
	note = {ADS Bibcode: 1992ApJ...401..206C},
	pages = {206},
}

@article{inoue_turbulence_2009,
	title = {Turbulence and Magnetic Field Amplification in Supernova Remnants: Interactions Between A Strong Shock Wave and Multi-Phase Interstellar Medium},
	volume = {695},
	issn = {0004-637X, 1538-4357},
	shorttitle = {{TURBULENCE} {AND} {MAGNETIC} {FIELD} {AMPLIFICATION} {IN} {SUPERNOVA} {REMNANTS}},
	url = {https://iopscience.iop.org/article/10.1088/0004-637X/695/2/825},
	doi = {10.1088/0004-637X/695/2/825},
	language = {en},
	number = {2},
	urldate = {2026-02-16},
	journal = {The Astrophysical Journal},
	author = {Inoue, Tsuyoshi and Yamazaki, Ryo and Inutsuka, Shu-ichiro},
	month = apr,
	year = {2009},
	pages = {825--833},
}

@article{lee_turbulence_2020,
	title = {Turbulence {Spectra} of {Electron} {Density} and {Magnetic} {Field} {Fluctuations} in the {Local} {Interstellar} {Medium}},
	volume = {904},
	issn = {0004-637X, 1538-4357},
	url = {https://iopscience.iop.org/article/10.3847/1538-4357/abba20},
	doi = {10.3847/1538-4357/abba20},
	language = {en},
	number = {1},
	urldate = {2026-02-16},
	journal = {The Astrophysical Journal},
	author = {Lee, K. H. and Lee, L. C.},
	month = nov,
	year = {2020},
	pages = {66},
}

@article{ressler_electron_2015,
	title = {Electron thermodynamics in {GRMHD} simulations of low-luminosity black hole accretion},
	volume = {454},
	issn = {0035-8711, 1365-2966},
	url = {https://academic.oup.com/mnras/article-lookup/doi/10.1093/mnras/stv2084},
	doi = {10.1093/mnras/stv2084},
	language = {en},
	number = {2},
	urldate = {2026-02-16},
	journal = {Monthly Notices of the Royal Astronomical Society},
	author = {Ressler, S. M. and Tchekhovskoy, A. and Quataert, E. and Chandra, M. and Gammie, C. F.},
	month = dec,
	year = {2015},
	pages = {1848--1870},
}

@article{event_horizon_telescope_collaboration_first_2022,
	title = {First {Sagittarius} {A}* {Event} {Horizon} {Telescope} {Results}. {V}. {Testing} {Astrophysical} {Models} of the {Galactic} {Center} {Black} {Hole}},
	volume = {930},
	issn = {2041-8205, 2041-8213},
	url = {https://iopscience.iop.org/article/10.3847/2041-8213/ac6672},
	doi = {10.3847/2041-8213/ac6672},
	language = {en},
	number = {2},
	urldate = {2026-02-16},
	journal = {The Astrophysical Journal Letters},
	author = {{Event Horizon Telescope Collaboration} and Akiyama, Kazunori and Alberdi, Antxon and Alef, Walter and Carlos Algaba, Juan and Anantua, Richard and Asada, Keiichi and Azulay, Rebecca and Bach, Uwe and Baczko, Anne-Kathrin and Ball, David and Baloković, Mislav and Barrett, John and Bauböck, Michi and Benson, Bradford A. and Bintley, Dan and Blackburn, Lindy and Blundell, Raymond and Bouman, Katherine L. and Bower, Geoffrey C. and Boyce, Hope and Bremer, Michael and Brinkerink, Christiaan D. and Brissenden, Roger and Britzen, Silke and Broderick, Avery E. and Broguiere, Dominique and Bronzwaer, Thomas and Bustamante, Sandra and Byun, Do-Young and Carlstrom, John E. and Ceccobello, Chiara and Chael, Andrew and Chan, Chi-kwan and Chatterjee, Koushik and Chatterjee, Shami and Chen, Ming-Tang and Chen 陈, Yongjun 永军 and Cheng, Xiaopeng and Cho, Ilje and Christian, Pierre and Conroy, Nicholas S. and Conway, John E. and Cordes, James M. and Crawford, Thomas M. and Crew, Geoffrey B. and Cruz-Osorio, Alejandro and Cui 崔, Yuzhu 玉竹 and Davelaar, Jordy and De Laurentis, Mariafelicia and Deane, Roger and Dempsey, Jessica and Desvignes, Gregory and Dexter, Jason and Dhruv, Vedant and Doeleman, Sheperd S. and Dougal, Sean and Dzib, Sergio A. and Eatough, Ralph P. and Emami, Razieh and Falcke, Heino and Farah, Joseph and Fish, Vincent L. and Fomalont, Ed and Ford, H. Alyson and Fraga-Encinas, Raquel and Freeman, William T. and Friberg, Per and Fromm, Christian M. and Fuentes, Antonio and Galison, Peter and Gammie, Charles F. and García, Roberto and Gentaz, Olivier and Georgiev, Boris and Goddi, Ciriaco and Gold, Roman and Gómez-Ruiz, Arturo I. and Gómez, José L. and Gu 顾, Minfeng 敏峰 and Gurwell, Mark and Hada, Kazuhiro and Haggard, Daryl and Haworth, Kari and Hecht, Michael H. and Hesper, Ronald and Heumann, Dirk and Ho 何, Luis C. 子山 and Ho, Paul and Honma, Mareki and Huang, Chih-Wei L. and Huang 黄, Lei 磊 and Hughes, David H. and Ikeda, Shiro and Violette Impellizzeri, C. M. and Inoue, Makoto and Issaoun, Sara and James, David J. and Jannuzi, Buell T. and Janssen, Michael and Jeter, Britton and Jiang 江, Wu 悟 and Jiménez-Rosales, Alejandra and Johnson, Michael D. and Jorstad, Svetlana and Joshi, Abhishek V. and Jung, Taehyun and Karami, Mansour and Karuppusamy, Ramesh and Kawashima, Tomohisa and Keating, Garrett K. and Kettenis, Mark and Kim, Dong-Jin and Kim, Jae-Young and Kim, Jongsoo and Kim, Junhan and Kino, Motoki and Koay, Jun Yi and Kocherlakota, Prashant and Kofuji, Yutaro and Koch, Patrick M. and Koyama, Shoko and Kramer, Carsten and Kramer, Michael and Krichbaum, Thomas P. and Kuo, Cheng-Yu and Bella, Noemi La and Lauer, Tod R. and Lee, Daeyoung and Lee, Sang-Sung and Leung, Po Kin and Levis, Aviad and Li 李, Zhiyuan 志远 and Lico, Rocco and Lindahl, Greg and Lindqvist, Michael and Lisakov, Mikhail and Liu 刘, Jun 俊 and Liu, Kuo and Liuzzo, Elisabetta and Lo, Wen-Ping and Lobanov, Andrei P. and Loinard, Laurent and Lonsdale, Colin J. and Lu 路, Ru-Sen 如森 and Mao 毛, Jirong 基荣 and Marchili, Nicola and Markoff, Sera and Marrone, Daniel P. and Marscher, Alan P. and Martí-Vidal, Iván and Matsushita, Satoki and Matthews, Lynn D. and Medeiros, Lia and Menten, Karl M. and Michalik, Daniel and Mizuno, Izumi and Mizuno, Yosuke and Moran, James M. and Moriyama, Kotaro and Moscibrodzka, Monika and Müller, Cornelia and Mus, Alejandro and Musoke, Gibwa and Myserlis, Ioannis and Nadolski, Andrew and Nagai, Hiroshi and Nagar, Neil M. and Nakamura, Masanori and Narayan, Ramesh and Narayanan, Gopal and Natarajan, Iniyan and Nathanail, Antonios and Navarro Fuentes, Santiago and Neilsen, Joey and Neri, Roberto and Ni, Chunchong and Noutsos, Aristeidis and Nowak, Michael A. and Oh, Junghwan and Okino, Hiroki and Olivares, Héctor and Ortiz-León, Gisela N. and Oyama, Tomoaki and Özel, Feryal and Palumbo, Daniel C. M. and Filippos Paraschos, Georgios and Park, Jongho and Parsons, Harriet and Patel, Nimesh and Pen, Ue-Li and Pesce, Dominic W. and Piétu, Vincent and Plambeck, Richard and PopStefanija, Aleksandar and Porth, Oliver and Pötzl, Felix M. and Prather, Ben and Preciado-López, Jorge A. and Psaltis, Dimitrios and Pu, Hung-Yi and Ramakrishnan, Venkatessh and Rao, Ramprasad and Rawlings, Mark G. and Raymond, Alexander W. and Rezzolla, Luciano and Ricarte, Angelo and Ripperda, Bart and Roelofs, Freek and Rogers, Alan and Ros, Eduardo and Romero-Cañizales, Cristina and Roshanineshat, Arash and Rottmann, Helge and Roy, Alan L. and Ruiz, Ignacio and Ruszczyk, Chet and Rygl, Kazi L. J. and Sánchez, Salvador and Sánchez-Argüelles, David and Sánchez-Portal, Miguel and Sasada, Mahito and Satapathy, Kaushik and Savolainen, Tuomas and Schloerb, F. Peter and Schonfeld, Jonathan and Schuster, Karl-Friedrich and Shao, Lijing and Shen 沈, Zhiqiang 志强 and Small, Des and Sohn, Bong Won and SooHoo, Jason and Souccar, Kamal and Sun 孙, He 赫 and Tazaki, Fumie and Tetarenko, Alexandra J. and Tiede, Paul and Tilanus, Remo P. J. and Titus, Michael and Torne, Pablo and Traianou, Efthalia and Trent, Tyler and Trippe, Sascha and Turk, Matthew and Van Bemmel, Ilse and Van Langevelde, Huib Jan and Van Rossum, Daniel R. and Vos, Jesse and Wagner, Jan and Ward-Thompson, Derek and Wardle, John and Weintroub, Jonathan and Wex, Norbert and Wharton, Robert and Wielgus, Maciek and Wiik, Kaj and Witzel, Gunther and Wondrak, Michael F. and Wong, George N. and Wu 吴, Qingwen 庆文 and Yamaguchi, Paul and Yoon, Doosoo and Young, André and Young, Ken and Younsi, Ziri and Yuan 袁, Feng 峰 and Yuan 袁, Ye-Fei 业飞 and Zensus, J. Anton and Zhang, Shuo and Zhao, Guang-Yao and Zhao 赵, Shan-Shan 杉杉 and Chan, Tin Lok and Qiu, Richard and Ressler, Sean and White, Chris},
	month = may,
	year = {2022},
	pages = {L16},
}

@article{tong_whistler_2019,
	title = {Whistler {Wave} {Generation} by {Halo} {Electrons} in the {Solar} {Wind}},
	volume = {870},
	issn = {2041-8205, 2041-8213},
	url = {https://iopscience.iop.org/article/10.3847/2041-8213/aaf734},
	doi = {10.3847/2041-8213/aaf734},
	language = {en},
	number = {1},
	urldate = {2026-02-16},
	journal = {The Astrophysical Journal Letters},
	author = {Tong, Yuguang and Vasko, Ivan Y. and Pulupa, Marc and Mozer, Forrest S. and Bale, Stuart D. and Artemyev, Anton V. and Krasnoselskikh, Vladimir},
	month = jan,
	year = {2019},
	pages = {L6},
}

@article{tong_statistical_2019,
	title = {Statistical {Study} of {Whistler} {Waves} in the {Solar} {Wind} at 1 au},
	volume = {878},
	issn = {0004-637X, 1538-4357},
	url = {https://iopscience.iop.org/article/10.3847/1538-4357/ab1f05},
	doi = {10.3847/1538-4357/ab1f05},
	language = {en},
	number = {1},
	urldate = {2026-02-16},
	journal = {The Astrophysical Journal},
	author = {Tong, Yuguang and Vasko, Ivan Y. and Artemyev, Anton V. and Bale, Stuart D. and Mozer, Forrest S.},
	month = jun,
	year = {2019},
	pages = {41},
}

@article{svenningsson_kinetic_2022,
	title = {Kinetic {Generation} of {Whistler} {Waves} in the {Turbulent} {Magnetosheath}},
	volume = {49},
	copyright = {All rights reserved},
	issn = {0094-8276, 1944-8007},
	url = {https://onlinelibrary.wiley.com/doi/10.1029/2022GL099065},
	doi = {10.1029/2022GL099065},
	language = {en},
	number = {15},
	urldate = {2023-01-20},
	journal = {Geophysical Research Letters},
	author = {Svenningsson, I. and Yordanova, E. and Cozzani, G. and Khotyaintsev, Yu. V. and André, M.},
	month = aug,
	year = {2022},
}

@article{breuillard_properties_2018,
	title = {The {Properties} of {Lion} {Roars} and {Electron} {Dynamics} in {Mirror} {Mode} {Waves} {Observed} by the {Magnetospheric} {MultiScale} {Mission}},
	volume = {123},
	issn = {2169-9380, 2169-9402},
	url = {https://onlinelibrary.wiley.com/doi/10.1002/2017JA024551},
	doi = {10.1002/2017JA024551},
	language = {en},
	number = {1},
	urldate = {2023-01-20},
	journal = {Journal of Geophysical Research: Space Physics},
	author = {Breuillard, H. and Le Contel, O. and Chust, T. and Berthomier, M. and Retino, A. and Turner, D. L. and Nakamura, R. and Baumjohann, W. and Cozzani, G. and Catapano, F. and Alexandrova, A. and Mirioni, L. and Graham, D. B. and Argall, M. R. and Fischer, D. and Wilder, F. D. and Gershman, D. J. and Varsani, A. and Lindqvist, P.‐A. and Khotyaintsev, Yu. V. and Marklund, G. and Ergun, R. E. and Goodrich, K. A. and Ahmadi, N. and Burch, J. L. and Torbert, R. B. and Needell, G. and Chutter, M. and Rau, D. and Dors, I. and Russell, C. T. and Magnes, W. and Strangeway, R. J. and Bromund, K. R. and Wei, H. and Plaschke, F. and Anderson, B. J. and Le, G. and Moore, T. E. and Giles, B. L. and Paterson, W. R. and Pollock, C. J. and Dorelli, J. C. and Avanov, L. A. and Saito, Y. and Lavraud, B. and Fuselier, S. A. and Mauk, B. H. and Cohen, I. J. and Fennell, J. F.},
	month = jan,
	year = {2018},
	pages = {93--103},
}

@article{dimmock_statistical_2015,
	title = {A {Statistical} {Study} of the {Dawn}‐{Dusk} {Asymmetry} of {Ion} {Temperature} {Anisotropy} and {Mirror} {Mode} {Occurrence} in the {Terrestrial} {Dayside} {Magnetosheath} {Using} {THEMIS} {Data}},
	volume = {120},
	issn = {2169-9380, 2169-9402},
	url = {https://onlinelibrary.wiley.com/doi/10.1002/2015JA021192},
	doi = {10.1002/2015JA021192},
	language = {en},
	number = {7},
	urldate = {2023-01-20},
	journal = {Journal of Geophysical Research: Space Physics},
	author = {Dimmock, A. P. and Osmane, A. and Pulkkinen, T. I. and Nykyri, K.},
	month = jul,
	year = {2015},
	pages = {5489--5503},
}

@article{gary_whistler_1994,
	title = {The {Whistler} {Heat} {Flux} {Instability}: {Threshold} {Conditions} in the {Solar} {Wind}},
	volume = {99},
	issn = {0148-0227},
	shorttitle = {The {Whistler} {Heat} {Flux} {Instability}},
	url = {http://doi.wiley.com/10.1029/94JA02067},
	doi = {10.1029/94JA02067},
	language = {en},
	number = {A12},
	urldate = {2023-01-20},
	journal = {Journal of Geophysical Research},
	author = {Gary, S. Peter and Scime, Earl E. and Phillips, John L. and Feldman, William C.},
	year = {1994},
	pages = {23391},
}

@article{gary_whistler_1996,
	title = {Whistler {Instability}: {Electron} {Anisotropy} {Upper} {Bound}},
	volume = {101},
	issn = {01480227},
	shorttitle = {Whistler {Instability}},
	url = {http://doi.wiley.com/10.1029/96JA00323},
	doi = {10.1029/96JA00323},
	language = {en},
	number = {A5},
	urldate = {2023-01-20},
	journal = {Journal of Geophysical Research: Space Physics},
	author = {Gary, S. Peter and Wang, Joseph},
	month = may,
	year = {1996},
	pages = {10749--10754},
}

@article{graham_nonmaxwellianity_2021,
	title = {Non‐{Maxwellianity} of {Electron} {Distributions} {Near} {Earth}'s {Magnetopause}},
	volume = {126},
	issn = {2169-9380, 2169-9402},
	url = {https://onlinelibrary.wiley.com/doi/10.1029/2021JA029260},
	doi = {10.1029/2021JA029260},
	language = {en},
	number = {10},
	urldate = {2023-01-20},
	journal = {Journal of Geophysical Research: Space Physics},
	author = {Graham, D. B. and Khotyaintsev, Yu V. and André, M. and Vaivads, A. and Chasapis, A. and Matthaeus, W. H. and Retinò, A. and Valentini, F. and Gershman, D. J.},
	month = oct,
	year = {2021},
}

@article{yordanova_electron_2016,
	title = {Electron {Scale} {Structures} and {Magnetic} {Reconnection} {Signatures} in the {Turbulent} {Magnetosheath}: {Current} {Sheets} in the {Magnetosheath}},
	volume = {43},
	issn = {00948276},
	shorttitle = {Electron {Scale} {Structures} and {Magnetic} {Reconnection} {Signatures} in the {Turbulent} {Magnetosheath}},
	url = {http://doi.wiley.com/10.1002/2016GL069191},
	doi = {10.1002/2016GL069191},
	language = {en},
	number = {12},
	urldate = {2023-01-23},
	journal = {Geophysical Research Letters},
	author = {Yordanova, E. and V{\"o}r{\"o}s, Z. and Varsani, A. and Graham, D. B. and Norgren, C. and Khotyaintsev, Yu. V. and Vaivads, A. and Eriksson, E. and Nakamura, R. and Lindqvist, P.-A. and Marklund, G. and Ergun, R. E. and Magnes, W. and Baumjohann, W. and Fischer, D. and Plaschke, F. and Narita, Y. and Russell, C. T. and Strangeway, R. J. and Le Contel, O. and Pollock, C. and Torbert, R. B. and Giles, B. J. and Burch, J. L. and Avanov, L. A. and Dorelli, J. C. and Gershman, D. J. and Paterson, W. R. and Lavraud, B. and Saito, Y.},
	month = jun,
	year = {2016},
	pages = {5969--5978},
}

@article{chen_evidence_2019,
	title = {Evidence for {Electron} {Landau} {Damping} in {Space} {Plasma} {Turbulence}},
	volume = {10},
	issn = {2041-1723},
	url = {https://www.nature.com/articles/s41467-019-08435-3},
	doi = {10.1038/s41467-019-08435-3},
	language = {en},
	number = {1},
	urldate = {2023-01-23},
	journal = {Nature Communications},
	author = {Chen, C. H. K. and Klein, K. G. and Howes, G. G.},
	month = feb,
	year = {2019},
	pages = {740},
}

@article{kitamura_observations_2020,
	title = {Observations of the {Source} {Region} of {Whistler} {Mode} {Waves} in {Magnetosheath} {Mirror} {Structures}},
	volume = {125},
	issn = {2169-9380, 2169-9402},
	url = {https://onlinelibrary.wiley.com/doi/abs/10.1029/2019JA027488},
	doi = {10.1029/2019JA027488},
	language = {en},
	number = {5},
	urldate = {2023-01-23},
	journal = {Journal of Geophysical Research: Space Physics},
	author = {Kitamura, N. and Omura, Y. and Nakamura, S. and Amano, T. and Boardsen, S. A. and Ahmadi, N. and Le Contel, O. and Lindqvist, P.‐A. and Ergun, R. E. and Saito, Y. and Yokota, S. and Gershman, D. J. and Paterson, W. R. and Pollock, C. J. and Giles, B. L. and Russell, C. T. and Strangeway, R. J. and Burch, J. L.},
	month = may,
	year = {2020},
}

@article{page_generation_2021,
	title = {Generation of {High}-frequency {Whistler} {Waves} in the {Earth}’s {Quasi}-perpendicular {Bow} {Shock}},
	volume = {919},
	issn = {2041-8205, 2041-8213},
	url = {https://iopscience.iop.org/article/10.3847/2041-8213/ac2748},
	doi = {10.3847/2041-8213/ac2748},
	language = {en},
	number = {2},
	urldate = {2023-01-23},
	journal = {The Astrophysical Journal Letters},
	author = {Page, Brent and Vasko, Ivan Y. and Artemyev, Anton V. and Bale, Stuart D.},
	month = oct,
	year = {2021},
	pages = {L17},
}

@article{huang_existence_2017,
	title = {On the {Existence} of the {Kolmogorov} {Inertial} {Range} in the {Terrestrial} {Magnetosheath} {Turbulence}},
	volume = {836},
	issn = {2041-8213},
	url = {https://iopscience.iop.org/article/10.3847/2041-8213/836/1/L10},
	doi = {10.3847/2041-8213/836/1/L10},
	language = {en},
	number = {1},
	urldate = {2023-01-23},
	journal = {The Astrophysical Journal},
	author = {Huang, S. Y. and Hadid, L. Z. and Sahraoui, F. and Yuan, Z. G. and Deng, X. H.},
	month = feb,
	year = {2017},
	pages = {L10},
}

@article{russell_magnetospheric_2016,
	title = {The {Magnetospheric} {Multiscale} {Magnetometers}},
	volume = {199},
	issn = {0038-6308, 1572-9672},
	url = {http://link.springer.com/10.1007/s11214-014-0057-3},
	doi = {10.1007/s11214-014-0057-3},
	language = {en},
	number = {1-4},
	urldate = {2023-01-23},
	journal = {Space Science Reviews},
	author = {Russell, C. T. and Anderson, B. J. and Baumjohann, W. and Bromund, K. R. and Dearborn, D. and Fischer, D. and Le, G. and Leinweber, H. K. and Leneman, D. and Magnes, W. and Means, J. D. and Moldwin, M. B. and Nakamura, R. and Pierce, D. and Plaschke, F. and Rowe, K. M. and Slavin, J. A. and Strangeway, R. J. and Torbert, R. and Hagen, C. and Jernej, I. and Valavanoglou, A. and Richter, I.},
	month = mar,
	year = {2016},
	pages = {189--256},
}

@article{burch_magnetospheric_2016,
	title = {Magnetospheric {Multiscale} {Overview} and {Science} {Objectives}},
	volume = {199},
	issn = {0038-6308, 1572-9672},
	url = {http://link.springer.com/10.1007/s11214-015-0164-9},
	doi = {10.1007/s11214-015-0164-9},
	language = {en},
	number = {1-4},
	urldate = {2023-01-23},
	journal = {Space Science Reviews},
	author = {Burch, J. L. and Moore, T. E. and Torbert, R. B. and Giles, B. L.},
	month = mar,
	year = {2016},
	pages = {5--21},
}

@article{pollock_fast_2016,
	title = {Fast {Plasma} {Investigation} for {Magnetospheric} {Multiscale}},
	volume = {199},
	issn = {0038-6308, 1572-9672},
	url = {http://link.springer.com/10.1007/s11214-016-0245-4},
	doi = {10.1007/s11214-016-0245-4},
	language = {en},
	number = {1-4},
	urldate = {2023-01-23},
	journal = {Space Science Reviews},
	author = {Pollock, C. and Moore, T. and Jacques, A. and Burch, J. and Gliese, U. and Saito, Y. and Omoto, T. and Avanov, L. and Barrie, A. and Coffey, V. and Dorelli, J. and Gershman, D. and Giles, B. and Rosnack, T. and Salo, C. and Yokota, S. and Adrian, M. and Aoustin, C. and Auletti, C. and Aung, S. and Bigio, V. and Cao, N. and Chandler, M. and Chornay, D. and Christian, K. and Clark, G. and Collinson, G. and Corris, T. and De Los Santos, A. and Devlin, R. and Diaz, T. and Dickerson, T. and Dickson, C. and Diekmann, A. and Diggs, F. and Duncan, C. and Figueroa-Vinas, A. and Firman, C. and Freeman, M. and Galassi, N. and Garcia, K. and Goodhart, G. and Guererro, D. and Hageman, J. and Hanley, J. and Hemminger, E. and Holland, M. and Hutchins, M. and James, T. and Jones, W. and Kreisler, S. and Kujawski, J. and Lavu, V. and Lobell, J. and LeCompte, E. and Lukemire, A. and MacDonald, E. and Mariano, A. and Mukai, T. and Narayanan, K. and Nguyan, Q. and Onizuka, M. and Paterson, W. and Persyn, S. and Piepgrass, B. and Cheney, F. and Rager, A. and Raghuram, T. and Ramil, A. and Reichenthal, L. and Rodriguez, H. and Rouzaud, J. and Rucker, A. and Saito, Y. and Samara, M. and Sauvaud, J.-A. and Schuster, D. and Shappirio, M. and Shelton, K. and Sher, D. and Smith, D. and Smith, K. and Smith, S. and Steinfeld, D. and Szymkiewicz, R. and Tanimoto, K. and Taylor, J. and Tucker, C. and Tull, K. and Uhl, A. and Vloet, J. and Walpole, P. and Weidner, S. and White, D. and Winkert, G. and Yeh, P.-S. and Zeuch, M.},
	month = mar,
	year = {2016},
	pages = {331--406},
}

@article{yordanova_current_2020,
	title = {Current {Sheet} {Statistics} in the {Magnetosheath}},
	volume = {7},
	issn = {2296-987X},
	url = {https://www.frontiersin.org/article/10.3389/fspas.2020.00002/full},
	doi = {10.3389/fspas.2020.00002},
	language = {en},
	urldate = {2023-01-23},
	journal = {Frontiers in Astronomy and Space Sciences},
	author = {Yordanova, Emiliya and V{\"o}r{\"o}s, Zoltán and Raptis, Savvas and Karlsson, Tomas},
	month = feb,
	year = {2020},
	pages = {2},
}

@article{yao_electron_2018,
	title = {Electron {Dynamics} in {Magnetosheath} {Mirror}‐{Mode} {Structures}},
	volume = {123},
	issn = {2169-9380, 2169-9402},
	url = {https://onlinelibrary.wiley.com/doi/10.1029/2018JA025607},
	doi = {10.1029/2018JA025607},
	language = {en},
	number = {7},
	urldate = {2023-01-23},
	journal = {Journal of Geophysical Research: Space Physics},
	author = {Yao, S. T. and Shi, Q. Q. and Liu, J. and Yao, Z. H. and Guo, R. L. and Ahmadi, N. and Degeling, A. W. and Zong, Q. G. and Wang, X. G. and Tian, A. M. and Russell, C. T. and Fu, H. S. and Pu, Z. Y. and Fu, S. Y. and Zhang, H. and Sun, W. J. and Li, L. and Xiao, C. J. and Feng, Y. Y. and Giles, B. L.},
	month = jul,
	year = {2018},
	pages = {5561--5570},
}

@article{stawarz_turbulence-driven_2022,
	title = {Turbulence-{Driven} {Magnetic} {Reconnection} and the {Magnetic} {Correlation} {Length}: {Observations} from {Magnetospheric} {Multiscale} in {Earth}'s {Magnetosheath}},
	volume = {29},
	issn = {1070-664X, 1089-7674},
	shorttitle = {Turbulence-{Driven} {Magnetic} {Reconnection} and the {Magnetic} {Correlation} {Length}},
	url = {https://aip.scitation.org/doi/10.1063/5.0071106},
	doi = {10.1063/5.0071106},
	language = {en},
	number = {1},
	urldate = {2023-01-23},
	journal = {Physics of Plasmas},
	author = {Stawarz, J. E. and Eastwood, J. P. and Phan, T. D. and Gingell, I. L. and Pyakurel, P. S. and Shay, M. A. and Robertson, S. L. and Russell, C. T. and Le Contel, O.},
	month = jan,
	year = {2022},
	pages = {012302},
}

@article{dimmock_statistical_2013,
	title = {The {Statistical} {Mapping} of {Magnetosheath} {Plasma} {Properties} {Based} on {THEMIS} {Measurements} in the {Magnetosheath} {Interplanetary} {Medium} {Reference} {Frame}: {Magnetosheath} {Statistical} {Mapping}},
	volume = {118},
	issn = {21699380},
	shorttitle = {The {Statistical} {Mapping} of {Magnetosheath} {Plasma} {Properties} {Based} on {Themis} {Measurements} in the {Magnetosheath} {Interplanetary} {Medium} {Reference} {Frame}},
	url = {http://doi.wiley.com/10.1002/jgra.50465},
	doi = {10.1002/jgra.50465},
	language = {en},
	number = {8},
	urldate = {2023-01-23},
	journal = {Journal of Geophysical Research: Space Physics},
	author = {Dimmock, A. P. and Nykyri, K.},
	month = aug,
	year = {2013},
	pages = {4963--4976},
}

@article{barrie_wavelet_2019,
	title = {Wavelet {Compression} {Performance} of {MMS}/{FPI} {Plasma} {Count} {Data} with {Plasma} {Environment}},
	volume = {6},
	issn = {2333-5084, 2333-5084},
	url = {https://onlinelibrary.wiley.com/doi/10.1029/2018EA000430},
	doi = {10.1029/2018EA000430},
	language = {en},
	number = {1},
	urldate = {2023-01-23},
	journal = {Earth and Space Science},
	author = {Barrie, A. C. and Smith, D. L. and Elkington, S. R. and Sternovsky, Z. and Silva, D. and Giles, B. L. and Schiff, C.},
	month = jan,
	year = {2019},
	pages = {116--135},
}

@article{verscharen_case_2022,
	title = {A {Case} for {Electron}-{Astrophysics}},
	volume = {54},
	issn = {0922-6435, 1572-9508},
	url = {https://link.springer.com/10.1007/s10686-021-09761-5},
	doi = {10.1007/s10686-021-09761-5},
	language = {en},
	number = {2-3},
	urldate = {2023-05-09},
	journal = {Experimental Astronomy},
	author = {Verscharen, Daniel and Wicks, Robert T. and Alexandrova, Olga and Bruno, Roberto and Burgess, David and Chen, Christopher H. K. and D’Amicis, Raffaella and De Keyser, Johan and De Wit, Thierry Dudok and Franci, Luca and He, Jiansen and Henri, Pierre and Kasahara, Satoshi and Khotyaintsev, Yuri and Klein, Kristopher G. and Lavraud, Benoit and Maruca, Bennett A. and Maksimovic, Milan and Plaschke, Ferdinand and Poedts, Stefaan and Reynolds, Christopher S. and Roberts, Owen and Sahraoui, Fouad and Saito, Shinji and Salem, Chadi S. and Saur, Joachim and Servidio, Sergio and Stawarz, Julia E. and Štverák, Štěpán and Told, Daniel},
	month = dec,
	year = {2022},
	pages = {473--519},
}

@article{verscharen_electron-driven_2022,
	title = {Electron-{Driven} {Instabilities} in the {Solar} {Wind}},
	volume = {9},
	issn = {2296-987X},
	url = {https://www.frontiersin.org/articles/10.3389/fspas.2022.951628/full},
	doi = {10.3389/fspas.2022.951628},
	language = {en},
	urldate = {2024-03-15},
	journal = {Frontiers in Astronomy and Space Sciences},
	author = {Verscharen, Daniel and Chandran, B. D. G. and Boella, E. and Halekas, J. and Innocenti, M. E. and Jagarlamudi, V. K. and Micera, A. and Pierrard, V. and Štverák, Š. and Vasko, I. Y. and Velli, M. and Whittlesey, P. L.},
	month = aug,
	year = {2022},
	pages = {951628},
}

@article{gary_electron_1999,
	title = {Electron {Heat} {Flux} {Constraints} in the {Solar} {Wind}},
	volume = {6},
	issn = {1070-664X, 1089-7674},
	url = {https://pubs.aip.org/pop/article/6/6/2607/263775/Electron-heat-flux-constraints-in-the-solar-wind},
	doi = {10.1063/1.873532},
	language = {en},
	number = {6},
	urldate = {2024-03-15},
	journal = {Physics of Plasmas},
	author = {Gary, S. Peter and Skoug, Ruth M. and Daughton, William},
	month = jun,
	year = {1999},
	pages = {2607--2612},
}

@article{feldman_electron_1983,
	title = {Electron {Velocity} {Distributions} {Near} the {Earth}'s {Bow} {Shock}},
	volume = {88},
	copyright = {http://onlinelibrary.wiley.com/termsAndConditions\#vor},
	issn = {0148-0227},
	url = {https://agupubs.onlinelibrary.wiley.com/doi/10.1029/JA088iA01p00096},
	doi = {10.1029/JA088iA01p00096},
	language = {en},
	number = {A1},
	urldate = {2024-04-29},
	journal = {Journal of Geophysical Research: Space Physics},
	author = {Feldman, W. C. and Anderson, R. C. and Bame, S. J. and Gary, S. P. and Gosling, J. T. and McComas, D. J. and Thomsen, M. F. and Paschmann, G. and Hoppe, M. M.},
	month = jan,
	year = {1983},
	pages = {96--110},
}

@article{micera_particle--cell_2020,
	title = {Particle-in-cell {Simulation} of {Whistler} {Heat}-flux {Instabilities} in the {Solar} {Wind}: {Heat}-flux {Regulation} and {Electron} {Halo} {Formation}},
	volume = {903},
	issn = {2041-8205, 2041-8213},
	shorttitle = {Particle-in-cell {Simulation} of {Whistler} {Heat}-flux {Instabilities} in the {Solar} {Wind}},
	url = {https://iopscience.iop.org/article/10.3847/2041-8213/abc0e8},
	doi = {10.3847/2041-8213/abc0e8},
	language = {en},
	number = {1},
	urldate = {2024-04-29},
	journal = {The Astrophysical Journal Letters},
	author = {Micera, A. and Zhukov, A. N. and López, R. A. and Innocenti, M. E. and Lazar, M. and Boella, E. and Lapenta, G.},
	month = nov,
	year = {2020},
	pages = {L23},
}

@article{cattell_parker_2022,
	title = {Parker {Solar} {Probe} {Evidence} for the {Absence} of {Whistlers} {Close} to the {Sun} to {Scatter} {Strahl} and to {Regulate} {Heat} {Flux}},
	volume = {924},
	issn = {2041-8205, 2041-8213},
	url = {https://iopscience.iop.org/article/10.3847/2041-8213/ac4015},
	doi = {10.3847/2041-8213/ac4015},
	language = {en},
	number = {2},
	urldate = {2024-04-29},
	journal = {The Astrophysical Journal Letters},
	author = {Cattell, C. and Breneman, A. and Dombeck, J. and Hanson, E. and Johnson, M. and Halekas, J. and Bale, S. D. and Dudok De Wit, T. and Goetz, K. and Goodrich, K. and Malaspina, D. and Pulupa, M. and Case, T. and Kasper, J. C. and Larson, D. and Stevens, M. and Whittlesey, P.},
	month = jan,
	year = {2022},
	pages = {L33},
}

@article{stawarz_properties_2019,
	title = {Properties of the {Turbulence} {Associated} with {Electron}-only {Magnetic} {Reconnection} in {Earth}’s {Magnetosheath}},
	volume = {877},
	issn = {2041-8205, 2041-8213},
	url = {https://iopscience.iop.org/article/10.3847/2041-8213/ab21c8},
	doi = {10.3847/2041-8213/ab21c8},
	language = {en},
	number = {2},
	urldate = {2024-04-29},
	journal = {The Astrophysical Journal Letters},
	author = {Stawarz, J. E. and Eastwood, J. P. and Phan, T. D. and Gingell, I. L. and Shay, M. A. and Burch, J. L. and Ergun, R. E. and Giles, B. L. and Gershman, D. J. and Contel, O. Le and Lindqvist, P.-A. and Russell, C. T. and Strangeway, R. J. and Torbert, R. B. and Argall, M. R. and Fischer, D. and Magnes, W. and Franci, L.},
	month = jun,
	year = {2019},
	pages = {L37},
}

@article{retino_situ_2007,
	title = {In {Situ} {Evidence} of {Magnetic} {Reconnection} in {Turbulent} {Plasma}},
	volume = {3},
	copyright = {http://www.springer.com/tdm},
	issn = {1745-2473, 1745-2481},
	url = {https://www.nature.com/articles/nphys574},
	doi = {10.1038/nphys574},
	language = {en},
	number = {4},
	urldate = {2024-07-31},
	journal = {Nature Physics},
	author = {Retinò, A. and Sundkvist, D. and Vaivads, A. and Mozer, F. and André, M. and Owen, C. J.},
	month = apr,
	year = {2007},
	pages = {235--238},
}

\end{document}